\documentclass[11pt]{article}
\usepackage{graphicx}
\usepackage{amsmath}
\usepackage{amssymb}
\usepackage{amsxtra}

\def\({\left(}
\def\){\right)}

\def\beq{\begin{equation}}
\def\eeq{\end{equation}}
\def\bea{\begin{eqnarray}}
\def\eea{\end{eqnarray}}
\title{BSM Cosmology from BSM Physics}
\author{Maxim Khlopov\\
Virtual Institute of Astroparticle physics\\
Université de Paris, CNRS, Astroparticule et Cosmologie,\\ F-75013 Paris, France\\ 
Center for Cosmoparticle physics ``Cosmion''\\ National Research Nuclear University MEPhI, 115409 Moscow, Russia\\
Research Institute of Physics, Sousthern Federal University,\\ Stachki 194, Rostov on Don 344090, Russia,\\ 
e-mail khlopov@apc.univ-paris7.fr}

\begin{document}
\maketitle

\begin{abstract}
Now Standard $\Lambda$CDM cosmology is based on physics Beyond the Standard Model (BSM), which in turn needs cosmological probes for its study. This vicious circle of problems can be resolved by methods of cosmoparticle physics, in which cosmological messengers of new physics provide sensitive model dependent probes for BSM physics. Such messengers, which are inevitably present in any BSM basis for now Standard cosmology, lead to deviations from the Standard cosmological paradigm. We give brief review of some possible cosmological features and messengers of BSM physics, which include balancing of baryon asymmetry and dark matter by sphaleron transitions, hadronic dark matter and exotic cosmic ray components, a solution for puzzles of direct dark matter searches in dark atom model, antimatter in baryon asymmetrical Universe as sensitive probe for models of inflation and baryosynthesis and its possible probe in AMS02 experiment, PBH and GW messengers of BSM models and phase transitions in early Universe. These aspects are discussed in the general framework of methods of cosmoparticle physics.

\end{abstract}


\noindent PACS: ... list of PACS codes

\section{Introduction}\label{s:intro}
The now Standard model of elementary particles appeals to its extension for recovery of its internal problems and/or embedding in the framework of unified description of the fundamental natural forces (see \cite{PPNP} for recent review). Such extensions are unavoidable in the fundamental physical basis for now Standard cosmological scenario, involving inflation, baryosynthesis and dark matter/energy \cite{Lindebook,Kolbbook,Rubakovbook1,Rubakovbook2,book,newBook,4,DMRev}. Probes for the BSM physics, underlying now standard cosmology, inevitably imply methods of cosmoparticle physics of cross disciplinary study of physical, astrophysical and cosmological signatures of new physics \cite{book,newBook}. Here we discuss some development of these methods presented at the XXIV Bled Workshop "What comes beyond the Standard models?" with special emphasis on the cosmological messengers of new physics, which can find positive evidence in the experimental data and thus acquire the meaning of signatures for the corresponding BSM models, specifying their classes and ranges of parameters.

If confirmed, such cosmological signatures should find explanation together with the basic elements of the modern cosmology. Therefore, the approach, which pretends on the unified description of Nature \cite{Norma,Norma2} should not only reproduce the Standard model of elementary particles and propose BSM features, which provide realistic description of inflation, baryosynthesis and dark matter, but should be in possession to confront possible signatures of new physics, which can go beyond the standard cosmological paradigm. 

Cosmological messengers of new physics can help to remove conspiracy of BSM physics, related with absence of its experimental evidence at the LHC, as well as conspiracy of BSM cosmology, reflected in concordance of the data of precision cosmology with now standard $\Lambda$CDM cosmological scenario \cite{ijmpd19}. Multimessenger cosmological probes can provide effective tool to study new physics at very high energy scale \cite{ICPPA20,ecu}. Signatures for new physics play especially important role in these studies. They can strongly reduce the possible class of BSM models and provide determination of their parameters with high precision.

We consider such possible signatures in the direct searches of dark matter (Section \ref{da}), in gravitational wave signals from coalescence of massive black holes and searches for antinuclear component of cosmic rays (Section \ref{pbh}. We specify open questions in their confrontation with the corresponding messengers of BSM physics. We discuss is the conclusive Section \ref{cpp} there signatures and their significance in the context of cosmoparticle physics of BSM physics and cosmology.
\section{Signatures of dark matter physics}
\subsection{Dark atom signature in direct dark matter searches}\label{da}
The highly significant positive result of underground direct dark matter search in DAMA/NaI and DAMA/LIBRA experiments \cite{rita} can hardly be explained in the framework of the Standard cosmological paradigm of Weakly Interacting Massive Particles (WIMP), taking into account negative results of direct WIMP searches by other groups (see \cite{PPNP} for review and references). Though these apparently contradicting results may be somehow explained by difference of experimental strategy and still admit WIMP interpretation, their non-WIMP interpretation seems much more probable, making the positive results of DAMA group the signature for dark atom nature of cosmological dark matter \cite{PPNP,ecu,I,kuksa}.

The idea that dark matter can be formed by stable particles with negative even charge $-2n$ bound in dark atoms with $n$ primordial helium nuclei can qualitatively explain negative results of direct WIMP searches based on the search for nuclear recoil from WIMP interaction \cite{PPNP,ecu,I}. Dark atom interaction with matter is determined by its nuclear interacting helium shell, so that cosmic dark atoms slow down in terrestrial matter and cannot cause significant recoil in underground experiments. However, in the matter of underground detector dark atoms can form low energy (few keV) bound states with nuclei of detector. The energy release in such binding possess annual modulation due to adjustment of local concentration of dark atoms to the incoming cosmic flux and can lead to the signal, detected in DAMA experiment.

Dark atoms represent strongly interacting asymmetric dark matter, since the corresponding models assume excess of $-2n$ charged particles over their $+2n$ antiparticles. Such excess can naturally be related with baryon asymmetry, if multiple charged particles possess electroweak charges and participate in electroweak sphaleron transitions. It is shown in \cite{arnab} that the excess of $-2n$ charged particles in model of Walking Technicolor (WTC) and $\bar U$ antiquarks (with the charge $-2/3$ of new stable generation can be balanced with baryon asymmetry and explain the observed dark matter density by dark atoms. The open question is whether such balance, which should also take place in the case stable 5th generation in the approach \cite{Norma2}, can lead to the sufficient excess of $\bar u_5$ antiquarks to implement the idea of dark atoms in this case.

Pending on the value of $-2n$ charge, multiple charged constituents of dark atoms form either Bohr-like $O$He atoms, binding -2 charged particles with primordial helium nucleus, or Thomson-like $X$He atoms for $n>1$. In the first case, double charged particles may be either composite, being formed by chromo-Coulomb binding in cluster $\bar U \bar U \bar U$ of stable antiquarks $\bar U$ with charge $-2/3$, or -2 stable technileptons or technibaryons. Heavy quark clusters have strongly suppressed interaction with nucleons, while techniparticles behave as leptons. It leads to rather peculiar properties of dark atom - they have a heavy lepton or lepton-like core and nuclear interacting helium shell, which determines their interaction with baryonic matter.

Though interaction of dark atoms with nuclei are determined by their helium shell and thus don't involve parameters of new physics, the problem needs development of special methods for its solution. The approach of \cite{timur}, assumed continuous extension of a classical three body problem to realistic quantum-mechanical description, taking into account finite size of interacting nuclei and helium shell, in order to reach self-consistent account for Coulomb repulsion and  nuclear attraction, which can lead to creation of a shallow potential well with low energy bound state in dark atom - nucleus interaction. The development of this approach is presented in \cite{timur2} for both Bohr-like and Thomson-like atoms. However, it becomes clear that probably the correct quantum-mechanical description should start from very beginning from quantum-mechanical nature of dark atom and numerical solutions for Schrodinger equation for dark-atom -nucleus quantum system. Development of self-consistent quantum-mechanical model of dark-atom interaction with nuclei and will make possible interpretation of the results \cite{rita} in terms of signature of dark atoms.
\subsection{Multimessenger probes for decaying dark matter}
Development of large scale experimental facilities like IceCube, HAWC, AUGER and LHAASO provides multimessenger astronomical probes for cosmological messengers of superhigh energy physics \cite{ketovSym}. The complex of LHAASO can provide unique measurement of ultra high energy photons, being in some cases most sensitive probe for existence of messengers of new physics at ultra-high energy scales. Superheavy decaying dark matter may be one of such messengers. Its decay products may contain ultrahigh energy neutrinos, photons, charged leptons and quarks.
\begin{figure}
    \begin{center}
        \includegraphics[scale=0.9]{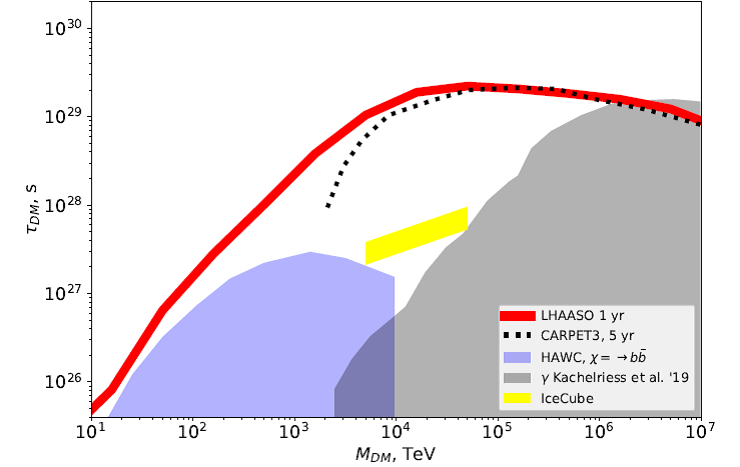}
        \caption{Multimessenger probes for decaying supermassive dark matter particles}
        \label{q}
    \end{center}
\end{figure}
Sensitivity of LHAASO for the measurement of dark matter decay time for DM decaying to quarks is demonstrated on Fig. \ref{q}, taken from \cite{andrea}. 
Yellow band shows the range of decay times for which DM decays give sizable contribution to the IceCube
neutrino signal \cite{neronov}. Blue and gray shaded regions show the existing bounds imposed by HAWC \cite{hawc} and
ultra-high-energy cosmic ray experiments \cite{ice}. and dashed curves are from the HAWC search of the DM
decay signal in the Fermi Bubble regions \cite{semikoz}. It makes possible to confront multimessenger cosmological probes with the data of multimessenger astronomy.
\section{Signatures for strong primordial inhomogeneities}\label{pbh}
\subsection{Massive PBHs}
Strong inhomogeneity of early Universe can lead to formation of primordial black holes (PBH). Such inhomogeneity may result from BSM physics at superhigh energy scales and thus even absence of positive evidence of PBH existence can provide important tool to probe allowed parameters of new physics at these scales \cite{ketovSym,PBHrev}. Formed within the cosmological horizon, which was small in the early Universe, it can seem that PBHs should have mass much smaller, than Solar mass $M_{\odot}$. However, the mechanisms of PBH formation can provide formation of PBHs with stellar mass, and even larger than stellar up to the seeds for Active Galactic nuclei (AGN) \cite{AGN,RubinCluster,dolgovPBH}.

LIGO/VIRGO detected gravitational wave signal from coalescence of black holes with masses exceeding the limit of pair instability  ($50 M_{\odot}$). Therefore black holes of such mass cannot be formed in the evolution of first stars. It has put forward the question on their primordial origin \cite{LV150,LV150apl} and may be considered as signature for BSM physics, underlying formation of such massive PBHs.

In the approach \cite{AGN} massive PBHs are formed in the collapse of closed walls originated from succession of phase transitions of breaking of U(1) symmetry and their mass is determined by the scale $f$ of spontaneous symmetry breaking at the inflationary stage and scale $\Lambda$ of successive explicit symmetry breaking. Therefore confirmation of primordial origin of massive PBHs would strongly narrow the choice of models of very early Universe
and its underlying physics.
\subsection{Cosmic antinuclei as probe for matter origin}
Baryon asymmetry of the Universe reflects absence of macroscopic antimatter in the amount comparable with baryonic matter within the observed Universe. Its origin is related with the mechanism of baryosynthesis, in which baryon excess is created in very early Universe. However, inhomogeneous baryosynthesis can lead not only to change of the value of baryon excess in different regions of space, but in the extreme case can change sign of this excess, giving rise to antimatter, produced in the same process, in which the baryonic matter was created \cite{CKSZ,DolgovAM,Dolgov2,Dolgov3,KRS2,AMS}. Antimatter domains should be sufficiently large to survive in matter surrounding and it implies also effect of inflation in addition to nonhomogeneous baryosynthesis. It means that the prediction of macroscopic antimatter, surviving to the present time, involves rather specific combination of necessary conditions and correspondingly specific choice of BSM model parameters. 

The choice of BSM model parameters determines the forms of macroscopic antimatter in our Galaxy. Antimatter domain can evolve in the way, similar to the baryonic matter and form antimatter globular cluster in our Galaxy\cite{AMS,GC}. The antibaryon density may be much higher, than the baryonic density and then specific ultra-dense antibaryon stars can be formed \cite{Blinnikov}, In any case, the predicted fraction of antihelium nuclei in cosmic rays from astrophysical sources is far below the sensitivity of AMS02 experiment, making positive results of cosmic antihelium signature of macroscopic antihelium in our Galaxy.

The possibility of confirmation of first indications to the antihelium events in AMS02 makes necessary to study in more details evolution of antibaryon domains in baryon asymmetrical universe \cite{orch,orch2} in the context of models of inhomogeneous baryosynthesis. It makes necessary to study expected composition and spectrum of cosmic antinuclei from antimatter globular cluster \cite{nastya}, as well as to consider more general question on propagation in galactic magnetic fields of antinuclei from local source in galactic halo \cite{nastya2}  
\section{Conclusions}\label{cpp}
There are some hints to new phenomena in the observational data \cite{nano,sunny,lev}. The deviations from the standard cosmological model may be related with the modified gravity \cite{mond,mond2}, leading beyond the Standard model of all the four fundamental interactions. Then one can expect additional types of polarization of gravitational waves \cite{sourav}. Such hints are not at such high significance level as the results of DAMA experiments \cite{rita}, but they can strongly extend the list of multimessenger probes of BSM physics.

Constraints on such exotic phenomena, as PBHs or antimatter in baryon asymmetrical Universe exclude rather narrow ranges of BSM model parameters.
Signatures for such phenomena make these ranges preferential, strongly reducing the class of BSM models and fixing their parameters with high precision. In the context of cosmoparticle physics, studying fundamental relationship of macro- and micro- worlds signatures for cosmological messengers of BSM physics acquires the meaning of precision measurement of parameters of fundamental structure of microworld with astronomical accuracy.  
\section*{Acknowledgements}
The work was supported by grant of Russian Science Foundation (Project No-18-12-00213-P).


\end{document}